\documentclass[twoside,11pt]{article}

\usepackage{cjp}
\usepackage{graphics}

\def\etal{\emph{et al.}}
\def\ie{\emph{i.e.}}
\def\insitu{\emph{in situ }}

\begin{document}

\title{The symmetry of phase-coherent thermopower oscillations in Andreev interferometers}

\author{Zhigang Jiang and Venkat Chandrasekhar}

\address{Department of Physics and Astronomy, Northwestern University, Evanston, IL 60208, USA}

\date{July 6, 2004}
\maketitle

\begin{abstract}
We study the thermopower of diffusive Andreev interferometers, which are hybrid loops with one normal-metal arm and one superconducting arm.  The thermopower oscillates as a function of the magnetic flux through the loop with a fundamental period corresponding to one flux quantum $\Phi_0=h/2e$. Unlike the electrical resistance oscillations and the thermal resistance oscillations, which are always symmetric with respect to the magnetic field, the symmetry of the thermopower oscillations can be either symmetric or antisymmetric depending on the geometry of the sample. We also observe that the symmetry of the thermopower oscillations is related to the distribution of the supercurrent between the normal-metal/superconductor interfaces. We compare our experimental results with recent theoretical work.
\end{abstract}


\section{Introduction}
When a temperature differential $\Delta T$ is established across a metallic sample and no electrical current is allowed to flow through it, an induced electrostatic potential differential $\Delta V$ will be set up across the sample. The thermopower $S$ is defined as the ratio of this voltage differential to the applied temperature differential $S\equiv\Delta V/\Delta T$.  For canonical metals, the thermopower is related to the energy-dependent conductivity $\sigma(\epsilon)$ by Mott's relation \cite{ashcroft}
\begin{equation}
S=-\frac{\pi^2}{3}\frac{k_B^2T}{e}\frac{\sigma'(\epsilon_F)}{\sigma(\epsilon_F)},
\label{eqn1}
\end{equation}
where $\sigma(\epsilon_F)$ the DC conductivity evaluated at the Fermi energy $\epsilon_F$ and $\sigma'(\epsilon_F)=\frac{\partial}{\partial \epsilon}\sigma(\epsilon)|_{\epsilon=\epsilon_F}$. In the framework of Fermi liquid theory, the thermopower stems from breaking of electron-hole symmetry, and arises from the second term in the Sommerfeld expansion \cite{ashcroft}. For a typical metal, this term is governed by a pre-factor ($k_BT/\epsilon_F$) and is usually very small.

In mesoscopic normal-metal/superconductor (NS) hybrid heterostructures, the properties of the electrons in the disordered normal metal are modified due to the proximity of the superconductor.  Mott's relation (\ref{eqn1}) is predicted to no longer be valid in this regime \cite{claughton, seviour}.  Experimentally, the thermopower of a normal metal in the proximity regime has been measured and is found to be much larger than the value estimated from Mott's relation \cite{eom1,eom2,eom3,dikin,parsons}. In addition, the electrons are phase coherent near the NS interfaces due to Andreev reflection \cite{andreev}. Hence, in doubly-connected structures such as Andreev interferometers, which are loops with one arm fabricated from a normal metal and the other arm from a superconductor, the thermopower oscillates as a function of the magnetic flux with a period corresponding to one flux quantum $\Phi_0=h/2e$.

\begin{figure}[!t]
\center{\includegraphics{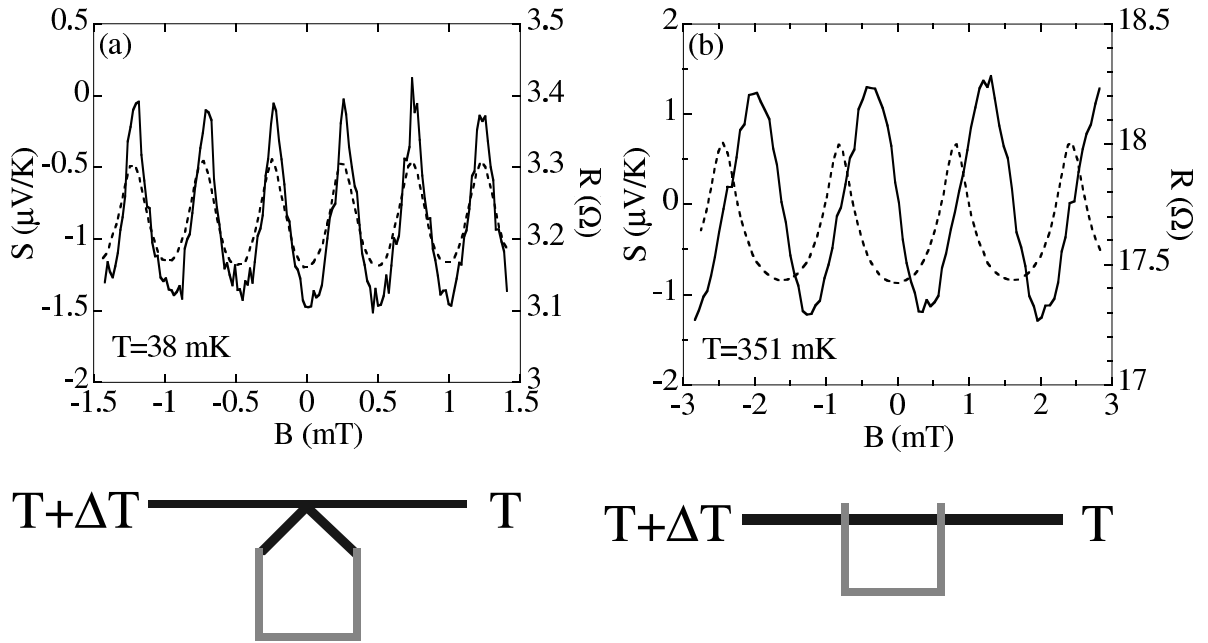}}
\caption{The thermopower (solid line) and the resistance (dashed line) oscillations as a function of magnetic field for two Andreev interferometers with different geometries: (a) `house' at $T=38$ mK and (b) `parallelogram' at $T=351$ mK. The schematics of Andreev interferometers are shown at the bottom of each panel.  The data are adapted from Ref. \cite{eom1}.}
\end{figure}

The most puzzling aspect of the experimental results on Andreev interferometers is that the symmetry of the thermopower oscillations with respect to the external magnetic field can be either symmetric or antisymmetric, depending on the geometry of the sample.  Figure 1, which has been adapted from Ref. \cite{eom1}, illustrates the dependence of the symmetry of the thermopower oscillations on the geometry of the sample.  Figure 1(a) shows data from an Andreev interferometer followed by the schematics of the sample.  Figure 1(b) shows data and schematics of another Andreev interferometer with different geometry. Following Ref. \cite{eom1}, we shall call these different sample geometries the `house' and the `parallelogram' interferometers respectively.  The dashed lines in Fig. 1 show the resistance of the interferometers as a function of applied magnetic field.  For both sample geometries, the resistance oscillates symmetrically with respect to the applied magnetic field.  The solid lines in Fig. 1 show the thermopower of the two interferometers as a function of magnetic field.  The thermopower of both interferometers oscillates periodically with magnetic field; however, while the oscillations for the `house' thermometer are symmetric with respect to magnetic field, the oscillations for the `parallelogram' interferometer are antisymmetric with respect to magnetic field.  Since these initial experiments, we have tried a number of different sample topologies.  All of these sample topologies, except the `house' topology, show a thermopower that is antisymmetric with respect to magnetic field.  In spite of considerable theoretical efforts, this dependence of the symmetry of the oscillations on the topology of the sample is not understood.    
 
In the following parts of this paper, we will first review some possible mechanisms that have been proposed to explain the experimental results, in particular, some recent theoretical work by Heikkil\"a's group \cite{heikkila}.  We will then report a new set of measurements  on samples where we can control the distribution of the supercurrents in the devices.  We find that the symmetry of the thermopower depends on the direction of the supercurrent, and the temperature dependence of the amplitude of the thermopower oscillations is determined by the correlation energy $E_c=\hbar D/L^2$, where $D$ is the electronic diffusion coefficient of the normal metal, and $L$ is the length of normal metal between the NS interfaces.

\section{Theoretical background}
The dependence of the electric current $I$ and thermal current $I^T$ on the voltage differential $\Delta V$ and temperature differential $\Delta T$ across a mesoscopic device can be expressed in terms of the two transport equations \cite{ashcroft}
\begin{equation}
I = G \Delta V + \eta \Delta T 
\label{eqn2}
\end{equation}
and
\begin{equation}
I^T = \zeta \Delta V + \kappa \Delta T. 
\label{eqn3}
\end{equation}
The thermopower is defined as the voltage differential generated by a temperature differential, under the condition that the electric current $I=0$, \ie, $S\equiv\Delta V/\Delta T$.  From Eqn. (\ref{eqn2}), this gives $S=-\eta/G$.  The off-diagonal thermoelectric coefficients $\eta$ and $\zeta$ are responsible for coupling $\Delta T$ to $I$, and $\Delta V$ to $I^T$.  For most canonical metals, such as the Au films used in our experiments, these coefficients are small, of order $k_B T/\epsilon_F$, resulting in the small thermopower seen in these systems.

For a normal metal coupled to a superconductor, one can write down equivalent transport equations in the quasiclassical approximation \cite{belzig,chandrasekhar}
\begin{equation}
\vec{j}(\vec{R},T)=eN_0D \int dE \; (M_{33}\partial_{\vec{R}}h_T + Q h_L + M_{03}\partial_{\vec{R}}h_L)
\label{eqn4}
\end{equation} 
and
\begin{equation}
\vec{j}_{th}(\vec{R},T)=N_0D \int dE \;E(M_{00}\partial_{\vec{R}}h_L + Q h_T + M_{30}\partial_{\vec{R}}h_T).
\label{eqn5}
\end{equation}
Here $h_L$ and $h_T$ are the so-called longitudinal and transverse distribution functions, which in equilibrium have the functional forms
\begin{equation}
h_{L,T}=\frac{1}{2}\left[\tanh\left(\frac{E+eV}{2 k_B T}\right) \pm \tanh\left(\frac{E-eV}{2 k_B T}\right)\right]
\label{eqn6}
\end{equation} 
Expansion of the derivative of the distribution function $h_T$ in the first term in the integrand of Eqn. (\ref{eqn4}) in terms of the voltage and temperature differentials $\Delta V$ and $\Delta T$ for a simple proximity-coupled normal metal gives a term that involves only $\Delta V$, and not $\Delta T$, \ie, there is no off-diagonal or thermoelectric term relating the electric current to $\Delta T$ arising from this term.  This is in agreement with the well-known fact that the derivation of the quasiclassical approximation assumes particle-hole symmetry, and hence throws out from the beginning the usual small thermoelectric effects found in a typical metal.  However, the third term involving $\partial_{\vec{R}} h_L$ in the integrand in Eqn. (\ref{eqn4}) will generate a term proportional to $\Delta T$, and hence a thermoelectric effect.  The factor multiplying $\partial_{\vec{R}} h_L$, $M_{03}$, depends on the imaginary part of the complex phase $\chi$ of the quasiclassical Green's function.  This term vanishes under most conditions.  One notable exception is when a normal current is converted to a supercurrent, for example, in the well-known case of charge imbalance regime in superconductors, where thermoelectric effects have been observed \cite{mamin}.  

The electric current also contains a term corresponding to the supercurrent, the second  term in the integrand in Eqn. (\ref{eqn4}).  Virtanen and Heikkil\"a (VH) \cite{heikkila}, following the earlier work by Seviour and Volkov \cite{seviour}, point out that this term may lead to a thermoelectric contribution in an Andreev interferometer.  To illustrate their concept, consider again the `parallelogram' interferometer. If the Josephson coupling between the two NS interfaces of the interferometer is strong, the application of a magnetic field will generate a diamagnetic supercurrent in the interferometer loop, which is of course antisymmetric in the applied magnetic flux.  If the temperatures of the two NS interfaces are not the same, the supercurrent coming out of one junction will not be the same as the current going in to the second junction, as the supercurrent is a strong function of temperature.  The excess current must go into the normal-metal side arms as a normal quasiparticle current .  Now the thermopower is measured under the condition that the total current through the sample vanishes.  A voltage must therefore develop across the sample that cancels the contribution due to the excess current.  This thermoelectric voltage will be antisymmetric in the applied magnetic flux, as the supercurrent is antisymmetric.  The amplitude of this thermal voltage is directly related to the resistance of the side arms (in the case of perfect NS contacts, the superconducting arm of the interferometer shorts out the contribution to the resistance of the normal arm); the larger the resistance, the larger the thermal voltage generated, and hence the larger the thermopower.

VH's analysis of the thermopower of the `parallelogram' interferometer requires the two NS interfaces to be at two different temperatures, so that there is an imbalance in the supercurrents entering the interfaces.  In the `parallelogram' configuration, where there is a thermal current along the normal arm of the interferometer, it is natural that such a difference in temperature exists.  For the `house' configuration, where there is no thermal current along the normal arm between two NS interfaces (the superconductors act as thermally insulating boundaries), one would assume that both NS interfaces should be at the same temperature, so the mechanism of VH would not generate a thermal voltage in this configuration.  Indeed, VH explicitly state in their paper that they cannot explain the symmetric thermopower observed in the experiments. The symmetric thermopower of the `house' Andreev interferometer indicates that there might be another effect that contributes to the thermopower in NS systems, which cannot be described within the quasiclassical approximation.

In addition, VH predict a nonmonotonic temperature dependence of the thermopower with a characteristic energy scale (where the thermopower reaches its maximum amplitude) set by the correlation energy $E_c$, since the thermopower is induced by the imbalance of the supercurrents and should have the same energy scale \cite{courtois}. This behavior is qualitatively in agreement with the experiments.

It should be noted that earlier, Kogan {\etal } discussed the influence of branch imbalance on the thermopower of NS systems \cite{kogan}.  The basic idea (to our understanding) is that the supercurrent in the superconducting portions of the device in the presence of a magnetic field may not be of the same magnitude as the supercurrent in the proximity-coupled normal metal, leading to conversion of the supercurrent to quasiparticle current in the vicinity of the NS interface, and hence the possibility of a branch imbalance (or an imbalance between electrons and holes).  This in turn leads to a thermoelectric voltage in the presence of a temperature gradient.  It is quite likely that this mechanism is in play in our system.  However, we do not know what might cause a difference in the magnitude of the supercurrents in the superconductor and the proximity coupled normal metal.  It seems that the maximum supercurrent that can circulate in the Andreev interferometer in both the superconducting and normal metal parts in response to an external magnetic field is determined by the critical supercurrent in the proximity coupled normal metal, which is much less than the critical supercurrent in the superconductor itself.  Hence, we are unsure of the origin of this predicted difference in magnitudes of the supercurrent.

\section{Experimental results}
In order to investigate the relation between the thermopower and the supercurrent in Andreev interferometers, a double-loop interferometer was fabricated, as shown in Fig. 2.    The Andreev interferometer is shown schematically in Fig. 2, to the right of the heater, which is a metallic film of 25 $\mu$m long and 1 $\mu$m wide.  The interferometer consists of an 8.5 $\mu$m long and 100 nm wide Au wire, which is connected to the heater on  one end, and a normal Au contact (labelled `2') on the other end.  The wire is connected above and below to two superconducting Al wires, forming two interferometer loops.  Around each interferometer loop, a superconducting Al thin film field coil was fabricated.  Magnetic flux could be coupled into each interferometer loop by sending a dc current into its field coil (the interferometer loops were separated by a distance of more than 25 $\mu$m, so that cross-coupling of the magnetic flux was minimized).  By varying the direction of the dc current, the flux coupled to both loops could be varied in phase (perpendicular to the plane of the substrate, and in the same direction), or out of phase (perpendicular to the plane of the substrate, but in opposite directions).   Other relevant device parameters are as follows:  Au film thickness, 50 nm; Al film thickness, 100 nm; low temperature (300 mK) resistivity of the Au film, $\rho_{Au} \sim1.5$ $\mu\Omega$cm, corresponding to a diffusion constant of $D_{Au}\sim 264$ cm$^2$/s.  In order to ensure good interfaces at the NS contacts, an \insitu Ar$^+$ plasma etch was used to clean the Au surface before the Al deposition. The transparency of the Au/Al interface was checked by an on-chip test sample, which had a resistance of 0.14 $\Omega$ for a 0.01 $\mu$m$^2$ area at room temperature.

\begin{figure}[!t]
\center{\includegraphics{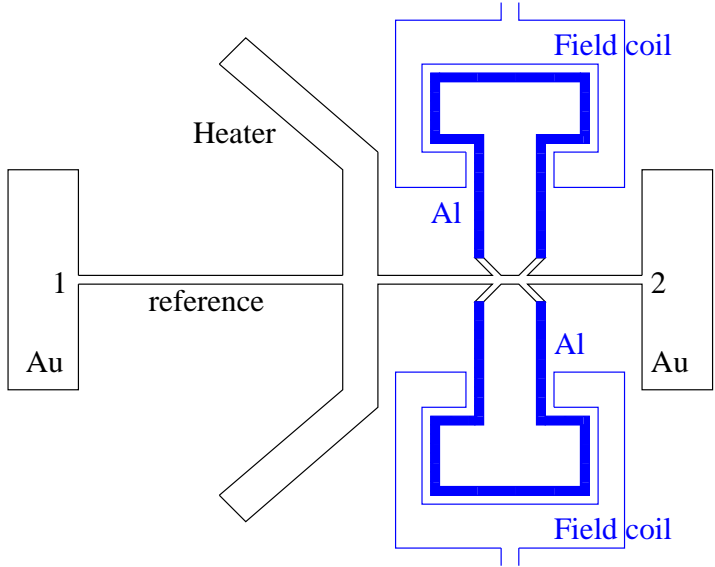}}
\caption{Schematic of the double-loop Andreev interferometer.}
\end{figure}

The technique for measuring the thermopower of NS devices has been described in many previous publications, so that here we shall concentrate only on the features specific to this experiment.  We use the second derivative technique described in Ref. \cite{dikin}.  An ac current $I$ of rms amplitude 5 $\mu$A with a frequency $f\sim$ 43 Hz was sent into the heater line, while $d^2V_{12}/dI^2$ is determined by measuring the ac voltage drop between contact pads `1' and `2' at a frequency of $2f$. (The normal-metal Au wire connected to contact pad `1' acts as a reference electrode for the thermopower measurements.)  Using the relation $d^2V_{12}/dI^2|_{I=0}=S_A(d^2T_h/dI^2|_{I=0})$ \cite{note}, where $S_A$ is the thermopower of Andreev interferometer and $T_h$ the local electron temperature at the `hot' end of the sample (the end connected to the heater line), and knowing $d^2T_h/dI^2$ is always symmetric with respect to the magnetic flux, one can obtain the symmetry of the thermopower oscillations directly from the symmetry of the measured $d^2V_{12}/dI^2$ \cite{dikin}.  Since we are interested only in the symmetry of the thermopower in this experiment, we do not measure the electron temperature using local proximity effect thermometers \cite{joe,jiang}, as we have done in previous experiments. 

The magnetic flux is applied locally by sending a dc current in series into the two field coils. As we have noted above, depending on the direction of the dc current, the fluxes coupled to the two loops can either be in phase, or out of phase. In the former case, assuming the device is perfectly symmetric, there will be no supercurrent along the path of the thermal current, as the supercurrent contributions from the two loops cancel each other, as shown in Fig. 3(a).  The in-phase flux configuration is therefore similar to the `house' geometry, in which no supercurrent flows along the path of the thermal current.  In the out-of-phase case (Fig. 3(b)), the two supercurrent contributions add, leading to a supercurrent that is twice the value for a single loop.  Since the supercurrent flows along the path of the temperature gradient, this configuration is similar to the `parallelogram' configuration.  

\begin{figure}[!t]
\center{\includegraphics{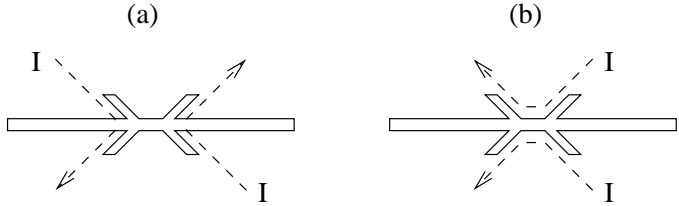}}
\caption{Preferred supercurrent distributions for (a) the in-phase flux configuration and (b) the out-of-phase flux configuration.}
\end{figure}

Figure 4 shows the thermopower and resistance of the double-loop Andreev interferometer as a function of the dc currents through the field coils, calibrated in units of the number of flux quanta through one loop.  The quantum of flux $\Phi_0$ through one loop was determined by sending the dc current through only one field coil, and measuring the resistance of the interferometer.  The resulting curves are shown as the dotted lines in Figs. 4(a) and 4(b).  For this sample, the amplitude of the resistance oscillations were only appreciable at higher temperatures, in the range of 0.75 to 1 K.  This is consistent with the reentrant behavior of the interferometer oscillations observed by other groups when the temperature is on the scale of the correlation temperature $E_c=\hbar D /L^2$ \cite{denhartog,stoof}.  Below a temperature scale on the order of $E_c/k_B$, the amplitude of the oscillations decreases.  Taking $L=500$ nm as the length between the NS interfaces, $E_c/k_B \sim$ 0.8 K  for this sample, hence the amplitude of the resistance oscillations decrease with decreasing temperature below this temperature range.  The  magnetoresistance data shown in this paper were all taken at a temperature of 0.93 K. 

The dashed lines in Figs. 4(a) and 4(b) show the resistance of the Andreev interferometer in the in-phase (Fig. 4(a)) and out-of-phase (Fig. 4(b)) configurations.  In both configurations, the resistance was found to be strongly hysteretic with magnetic flux, with the hysteresis increasing at lower temperatures.  This is consistent with a strong Josephson coupling between the NS interfaces, leading to a circulating supercurrent in response to the applied magnetic flux.  Consequently, we have only plotted the resistance for one direction of the sweep in both Figs. 4(a) and 4(b).  In the out-of-phase case (Fig. 4(a)), two periods can be discerned; the first corresponds to the period observed with the field applied only through one field coil (and hence corresponds to a flux quantum through only one loop), and a second smaller oscillation whose period is half that, corresponding to one flux quantum through both loops.  In the in-phase case, only oscillations of with period corresponding to a flux through one loop are observed.  At this point, we are not sure about the origin of this difference.  It should also be noted that the resistance oscillations are always \textit{symmetric} with respect to applied magnetic flux (the small offset seen in the data is most likely due to the Earth's magnetic field, since the area of the interferometer loops is large).     

The solid curves in Figs. 4(a) and 4(b) show the thermopower as a function of the dc current applied through both field coils, in the in-phase (Fig. 4(a)) and out-of-phase (Fig. 4(b)) configurations.  As with the electrical resistance, the amplitude of the thermopower oscillations decreased drastically at low temperatures, consistent with the energy scale being set again by $E_c$.  Consequently, these thermopower data were taken at a temperature of 0.79 K.  Although the amplitudes of the thermopower oscillations in the two configurations are approximately the same, the shape of the curves is quite different.  The oscillation waveform for the out-of-phase configuration is quite non-sinusoidal; furthermore, comparing the thermopower curve to the resistance curves, it can be seen that the thermopower in the out-of-phase configuration is antisymmetric with respect to the flux.  As we pointed out earlier, this configuration is similar to the `parallelogram' interferometer (which also shows an antisymmetric thermopower), in that there is a supercurrent in the path of the thermal current.  In contrast, the thermopower curve for the in-phase configuration shown in Fig. 4(a) is more sinusoidal, and it is symmetric with respect to the applied flux.  It should be emphasized that these two thermopower curves were taken from the same device, merely by changing how the flux (and hence the supercurrents) are distributed in the sample.  This shows that the symmetry of the thermpower is intimately related to whether or not supercurrent flows along the path of the thermal current.  However, at present, we do not know how the supercurrent couples to the thermopower.

\begin{figure}[!t]
\center{\includegraphics{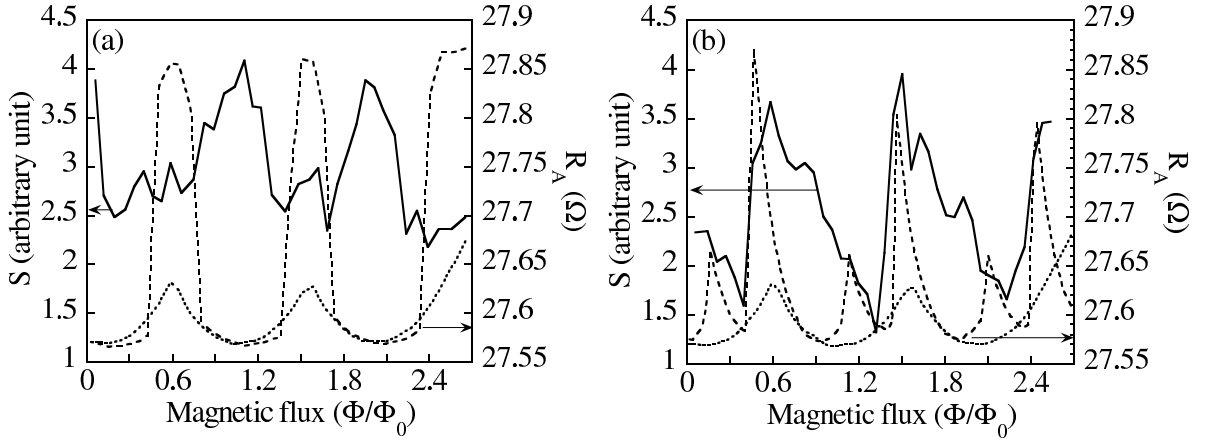}}
\caption{Thermopower (solid line) and resistance (dashed line) oscillations as a function of the dc currents through the field coils, calibrated in units of the number of flux quanta through one loop: (a) the in-phase configuration and (b) the out-of-phase configuration. The dotted line represents the resistance oscillations measured as sending a dc current through only one field coil. The thermopower is measured at $T=0.79$ K; the resistance is measured at $T=0.93$ K.}
\end{figure}

\section{Summary}
We find that phase-dependent thermopower oscillations in a proximity-coupled normal metal system are closely related to the supercurrent in such device, as predicted by recent theoretical work. The symmetry of the thermopower oscillations can be either symmetric or antisymmetric depending on the distribution of the supercurrent.   However, the detailed mechanism of the coupling of the thermopower to the supercurrent still needs to be determined.  We also observe that the amplitude of thermopower oscillations is related to the correlation energy $E_c$ of the system.  As was found before, the amplitude of the thermopower oscillations shows a non-monotonic dependence on the temperature, showing a maximum at some intermediate temperature $T$.  The experiments discussed here suggest that this temperature is related to the correlation energy of the sample, $T_{max} \sim E_c/k_B$.

\section*{Acknowledgment}
This work is supported by the NSF under grant number DMR-0201530.

% ---- Bibliography ----


\begin{thebibliography}{50}
\bibitem{note}The thermopower of the reference arm is small and does not vary as a function of the magnetic flux, so we neglect it in our analysis.
\end{thebibliography}
\end{document}